\documentclass [aps,12pt,aps,a4paper,prb]{revtex4}
\usepackage{graphicx}  
\usepackage{dcolumn}   
\usepackage{bm}        
\usepackage{amssymb}   

\begin{document}



\title{Counterion adsorption on flexible polyelectrolytes: comparison of theories}
\author{{\bf Rajeev Kumar, Arindam Kundagrami and M.Muthukumar} \footnote[1]{To whom any correspondence should be addressed, Email : muthu@polysci.umass.edu}}

\affiliation{\it Dept. of Polymer Science \& Engineering, Materials Research Science \& Engineering Center,\\
 University of Massachusetts,
 Amherst, MA-01003, USA.}
\date{\today}

\begin{abstract}
Counterion adsorption on a flexible polyelectrolyte chain in a
spherical cavity is considered by taking a ``permuted'' charge distribution on the chain
so that the ``adsorbed'' counterions are allowed to move along the backbone.
We compute the degree of ionization by using self-consistent field theory (SCFT) and
compare with the previously developed variational theory.
Analysis of various contributions to the free energy in both theories reveals that
the equilibrium degree of ionization is attained mainly as an interplay of the adsorption
energy of counterions on the backbone, the translational entropy of the small
ions, and their correlated density fluctuations. Degree of ionization computed from SCFT is significantly lower than
that from the variational formalism. The difference is entirely due to
the density fluctuations of the small ions in the system, which are
accounted for in the variational procedure. When these fluctuations
are deliberately suppressed in the truncated variational procedure,
there emerges a remarkable quantitative agreement in the various
contributing factors to the equilibrium degree of ionization, in spite
of the fundamental differences in the approximations and computational
procedures used in these two schemes. Furthermore, it is found that
the total free energies from the truncated variational procedure and
the SCFT are in quantitative agreement at low monomer densities and
differ from each other at higher monomer densities. The disagreement
at higher monomer densities is due to the inability of the variational
calculation to accurately compute the solvent entropy at higher
concentrations. A comparison of electrostatic energies (which are
relatively small) reveals
that the Debye-H\"{u}ckel estimate used in the variational theory is an
overestimation
of electrostatic energy as compared to the Poisson-Boltzmann estimate.
Nevertheless, since the significant effects from density fluctuations
of small ions are not captured by the SCFT, and due to the close
agreement between SCFT and the other contributing factors in the more
transparent variational procedure, the latter is a better
computational tool for obtaining the degree of ionization.

\end{abstract}

\maketitle

\section{Introduction}
Counterion adsorption in polyelectrolyte solutions is one of the fundamental problems
in polyelectrolyte physics, which has been a topic of extensive research\cite{rice61,manning69,dautzenberg,schmidtreview,monica95,stigter95,prabhu05,klein98,pincus98,kuhn05,liverpool06,
kremer95,winkler98,liu02,liughosh03,doby03,doby06,luijten06,yethiraj06,beer97,colby02,huber03,prabhu04,
hoagland04,essafi05,muthu04,grosberg02,arindam08} for decades. Historically\cite{rice61,manning69,dautzenberg,schmidtreview}, the counterion adsorption
in polyelectrolyte
solutions was described as an analog of ion-pairing in simple electrolyte solutions, and
deviations from the Debye-H\"{u}ckel limiting laws in colligative properties of polyelectrolyte solutions in dilute concentration regime had been attributed to counterion adsorption. Ignoring interactions among different chains in dilute solutions, modeling a polyelectrolyte chain by an infinite rod and using bulk dielectric constant of the medium for describing electrostatic interactions, counterion adsoprtion (or ``Manning condensation'')
was predicted to be a result of a singularity in the partition function arising due to a singular electrostatic potential near the rod. In particular, it was predicted that for monovalent monomers and counterions if $l_{B}/b > 1$, where $l_{B}$ is Bjerrum length ($=e^2/4\pi\epsilon_0 \epsilon k_B T$, $e$ being the electronic charge, $\epsilon_0$ being the dielectric constant of the vacuum, $\epsilon$ being the dielectric constant of the medium and $k_B T$ being the Boltzmann constant times temperature) and $b$ is the charge spacing on the backbone of the chain, then some of the counterions from the solution adsorb on the chain until $l_{B}/b = 1$. On the other hand, if $l_{B}/b<1$, the Debye-H\"{u}ckel approximation (i.e., the linearization of Poisson-Boltzmann) can be used to describe the electrostatics, and there is no adsorption.
In other words, counterion adsorption was described as a kind of phase transition
 (such as the condensation of vapors ) driven by electrostatic energy only.

Over the years, simulations\cite{kremer95,winkler98,liu02,liughosh03,doby03,doby06,luijten06,yethiraj06}
and experiments\cite{beer97,colby02,huber03,prabhu04,hoagland04,essafi05,prabhu05} have provided an insight into the counterion adsorption mechanism in polyelectrolyte solutions. Most importantly, it was revealed that a polyelectrolyte chain never attains a perfect rod conformation, even in salt-free solutions where the rod conformation is usually expected. Also, it has been shown that the effect of ion-pairing\cite{winkler98,hoagland04} on the backbone and dielectric constant\cite{essafi05} has to be considered in order to treat a realistic polyelectrolyte chain. Furthermore, simulations\cite{liu02,liughosh03} reveal that the counterions retain their translational degrees of freedom along the backbone of the chain after being adsorbed.

These insights obtained from the simulations and experiments have led to a number of
theoretical descriptions\cite{monica95,stigter95,klein98,pincus98,muthu04,kuhn05,liverpool06,
grosberg02,arindam08} of the counterion adsorption in polyelectrolyte solutions (containing
monovalent or multivalent salts) over the last two decades. For an isolated, \textit{single} polyelectrolyte chain,
it was shown that there is no singularity in the partition function when the polyelectrolyte chain is modeled as a flexible one\cite{muthu04}. An important prediction of the theory\cite{muthu04} emphasized the role played by the
dielectric mismatch between the local environment of the backbone and the bulk solution in driving counterion adsorption, and that had not been highlighted earlier in the literature.
In particular, it was shown that counterion adsorption arises as an interplay of adsorption energy and translational entropy of ions. Entropically, higher degree of ionization, an effect which is opposed by the lowering of electrostatic energy by the formation of ion pairs at the backbone of the chain, is preferred. To carry out the calculations analytically, the Debye-H\"{u}ckel potential was used to describe the electrostatic interactions between charged species, and a novel variational method was used to express the chain conformational entropy and excluded
 volume effects. Finally, the complete free energy of the system (consisting of the chain, the counterions, and the solvent) was simultaneously minimized in terms of the degree of counterion adsorption and the size of the polyelectrolyte chain to self-consistently obtain the equilibrium values of the respective variables. The theory is in qualitative agreement with the known simulation results. Recently, this single chain theory has been extended to describe the competitive counterion adsorption\cite{arindam08} phenomenon in the presence of multivalent and
monovalent counterions. Despite the qualitative agreements to simulations and experiments (especially, regarding the phenomena of charge reversal and reentrant transition), the approximations used in the theory have not rigorously been assessed so far.

In this work, we consider the counterion adsorption on a flexible polyelectrolyte chain in a spherical cavity in the presence of monovalent salt. Physically, such a situation is realized in extremely dilute polyelectrolyte solutions, where inter-chain interactions can be ignored safely and a finite volume can be carved out for each chain, and in pores confining polyelectrolyte chains, depending on the ratio of the cavity size to the radius of gyration of the polymer. We use the self-consistent field theory (SCFT) to compute the equilibrium degree of counterion adsorption. SCFT computes the free energy of the system by summing over all possible conformations of the chain, and hence, provides a more accurate description of the system (in fact, it provides the exact free energy at the mean-field level) compared to the variational formalism. Also, as the electrostatics in SCFT is treated at full, non-linear Poisson-Boltzmann level, we can assess the validity of the Debye-H\"{u}ckel potential to describe the electrostatic energy in the variational formalism.
Although SCFT provides an accurate and clear picture, it is computationally expensive to calculate
the degree of ionization due to a vast parameter space in the case of polyelectrolytes.
On the other hand, the variational theory put forward by Muthukumar\cite{muthu04} is transparent,
analytically tractable (to some extent), and very inexpensive
in terms of the computational needs. Aim of this study is to provide a simple, accurate, and easy-to-use method
to compute the degree of ionization and assess the approximations used in the variational formalism.

This paper is organized as follows: the theoretical formalisms are presented in Sec. ~\ref{sec:theory}; calculated results and conclusions are presented in Sec. ~\ref{sec:results} and ~\ref{sec:conclusions}, respectively.

\section{Comparison of theories : SCFT and variational formalism}
\setcounter {equation} {0} \label{sec:theory}
We consider a single flexible polyelectrolyte chain of
total $N$ Kuhn segments, each with length $l$, confined in a spherical
cavity of volume $\Omega = 4 \pi R^{3}/3$. The polyelectrolyte chain is represented
as a continuous curve of length $Nl$, and an arc length variable $t$
is used to represent any segment along the backbone so that $t \in [0,Nl]$. Now, we assume that
the chain (negatively charged) is surrounded by $n_{c}$ monovalent
counterions (positively charged) released by the chain along with
$n_{\gamma}$ ions of species $\gamma \,(= +,-)$ coming from added salt so that the
whole system is globally electroneutral.
Let $Z_{j}$ be the valency (with sign) of the charged species of type $j$,
and $n_{s}$  be the number of solvent molecules (satisfying
the incompressibility constraint after assuming the small ions to be pointlike) present
in the cavity. For simplicity, we have taken the volume of
a solvent molecule ($v_{s}$) to be equal to the volume
of the monomer (i.e., $v_{s}\equiv l^{3}$).
Subscripts $p,s,c,+$ and $-$ are used to represent
monomer, solvent , counterion from  polyelectrolyte, positive and negative salt ions, respectively.

In order to study counterion adsorption, we use the so-called ``two-state'' model for the
counterions so that there are two populations of counterions in the system. One population of the counterions
is free to enjoy the available volume (called the ``free'' counterions) and the other population is
 ``adsorbed'' on the backbone. However, the adsorbed counterions are allowed to enjoy
 translational degrees of freedom along the backbone,
 maintaining a total charge of $e f NZ_{p}$ on the chain, where $e$
is the electronic charge and $f$ is the degree of ionization of the chain (i.e., there are $-(1-f)NZ_{p}/Z_{c}$
``adsorbed'' counterions on the chain). In the literature, this kind of charge distribution has been referred to as
a ``permuted'' charge distribution\cite{orland98}.

For a particular set of parameters, we compute the free energy of the system comprising of the single chain, its counterions, the salt ions, and the solvent as a function of the degree of ionization ($f$). As mentioned before,
we compute the free energy using two different computational frameworks: SCFT\cite{fredbook,rajeev08} and the variational\cite{muthu04,muthuedwards,muthu87} formalism.
In both the formalisms, we ignore the electrostatic interactions between solvent molecules and
the small ions, and model the dielectric constant ($\epsilon$) of the medium to be independent of
temperature ($T$) to extract energy and entropy of the system. Also, for comparison purposes, we divide the
free energy into a mean field part and an additional part, which goes beyond the mean field theory.
Mean field part is further divided into
the contributions coming from the ``adsorbed'' counterions ($F_{a}^{\star}$) and the ``free'' ions, and from the
chain entropy etc., ($F_{f}^{\star}$). All contributions are properly identified(or subdivided into) as the enthalpic or(and) entropic parts. In all of what follows, the superscript $\star$ represents the mean field part. A brief description of
the derivation for the two formalisms is presented in appendices A and B,
and the original references\cite{muthu04,rajeev08} may be consulted for details.

To start with, we note that the contributions coming from ``adsorbed'' counterions are the same in both formalisms, and are given by
\begin{eqnarray}
F^{\star}_{a} &=& E_{a} - TS_{a}, \label{eq:free_adsorbed}\\
E_{a} &=& - (1-f)N\delta l_{B}/l, \label{eq:ea}\\
-TS_{a} &=& N\left[ f\log f + (1-f)\log (1-f)\right]\label{eq:sa},
\end{eqnarray}
so that $E_{a}$ is the electrostatic binding energy of the ion-pairs formed on the polymer
backbone due to the adsorption of ions
and $S_{a}$ is the translational entropy of the ``adsorbed'' counterions along the backbone. Parameter
$\delta = \epsilon l/\epsilon_{l}d$, reflects the deviation of the dielectric constant at the local
environment of the chain ($\epsilon_{l}$) from the bulk value ($\epsilon$), and $d$ represents the length of the dipole formed due to ion-pairing.

\subsection{Self-Consistent Field Theory}\label{sec:scft}
Although $F^{\star}_{a}$ is the same in both formalisms,
other contributions involving the ``free '' ions, the chain entropy etc. (i.e., $F^{\star}_{f}$) differ from
each other significantly in terms of computational details. In SCFT, $F^{\star}_{f}$ is computed
after solving for fields experienced by different components in the system,
which arise as a result of interactions of a particular component with the others.
For a single polyelectrolyte chain in a spherical cavity, each charged component (monomers and small ions)
experiences a dimensionless field $\psi$ (in units of $k_{B}T/e$), which is given by the solution
of the Poisson-Boltzmann equation
\begin{eqnarray}
\bigtriangledown_{\mathbf{r}}^{2}\psi(\mathbf{r}) &=& - 4 \pi l_{B} \rho_{e}(\mathbf{r}), \label{eq:saddle_pb}
\end{eqnarray}
where $\rho_{e}(\mathbf{r}) = \sum_{j = c,+,-} Z_{j}\rho_{j}(\mathbf{r}) + Z_{p}f \rho_{p}(\mathbf{r})$ is the local
charge density and $\rho_{\beta}(\mathbf{r})$ is the collective number density of
species of type $\beta = p,c,+,-$. Collective number densities for small ions are given by the Boltzmann
distribution with the prefactor determined by the constraint that the number of small ions are fixed in the system. Explicitly,

\begin{eqnarray}
\rho_{j}(\mathbf{r})&=& \frac{n_{j}\exp\left [- Z_{j}\psi(\mathbf{r}) \right]}{\int d\mathbf{r}\exp\left [- Z_{j}\psi(\mathbf{r}) \right]} \label{eq:den_small}
\end{eqnarray}
for $j=c,+,-$. In Eq. (~\ref{eq:saddle_pb}), $\rho_p(\mathbf{r})$ is the monomer density, which
is related to the probability of finding a particular monomer (described by the contour variable $t$) at a particular location $\mathbf{r}$, when the starting end of the chain can be anywhere in space (say, $q(\mathbf{r},t)$), by the relation
\begin{eqnarray}
\rho_{p}(\mathbf{r})&=&  \frac{\int_{0}^{N} dt \, q(\mathbf{r},t) q(\mathbf{r},N-t)}{\int d\mathbf{r}q(\mathbf{r},N)}. \label{eq:den_eq}
\end{eqnarray}
Furthermore, it can be shown that $q(\mathbf{r},t)$ satisfies the modified diffusion equation\cite{edwards65,fredbook}
\begin{eqnarray}
\frac{\partial q(\mathbf{r},t) }{\partial t} &=& \left [\frac{l^{2}}{6}\bigtriangledown_{\mathbf{r}}^{2}- \left \{ Z_{p}f \psi(\mathbf{r}) + w_{p}(\mathbf{r})\right \} \right ]q(\mathbf{r},t), \quad t\in (0,N), \label{eq:difreal}
\end{eqnarray}
$w_p$ being the field experienced by monomers due to non-electrostatic interactions.
At the saddle point, it is given by
\begin{eqnarray}
w_{p}(\mathbf{r}) &=& \chi_{ps}l^{3}\rho_{s}(\mathbf{r}) + \eta(\mathbf{r}),\label{eq:saddle_exp1}
\end{eqnarray}
where $\chi_{ps}$ is the dimensionless Flory's chi parameter
and $\eta(\mathbf{r})$ is the Lagrange's multiplier to enforce
the incompressibility constraint
\begin{eqnarray}
\rho_{p}(\mathbf{r}) &+& \rho_{s}(\mathbf{r}) = \rho_{0} \label{eq:saddle_incomp}
\end{eqnarray}
at all points in the system. Here, $\rho_{0}$ is the total number density
i.e., $\rho_{0} = (N + n_{s})/\Omega = 1/l^{3}$.
Finally, $\rho_s(\mathbf{r})$ is the collective number density of solvent molecules
given by the Boltzmann distribution in terms of the field experienced by a solvent molecule ($w_s$). Within
the saddle-point approximation, $\rho_s$ and $w_s$ are related by
\begin{eqnarray}
\rho_{s}(\mathbf{r})&=& \frac{n_{s}\exp\left [-w_{s}(\mathbf{r}) \right]}{\int d\mathbf{r}\exp\left [-w_{s}(\mathbf{r}) \right]},\label{eq:den_solv}\\
w_{s}(\mathbf{r}) &=& \chi_{ps}l^{3}\rho_{p}(\mathbf{r}) + \eta(\mathbf{r}).\label{eq:field_solv}
\end{eqnarray}
 These equations also close the loop of self-consistent equations, and Eqs.(~\ref{eq:saddle_pb}-~\ref{eq:field_solv}) form the set of coupled non-linear equations that
defines the system.

After solving these equations for fields (and in turn, for densities), the free energy at the saddle point, $F^{\star}_{f}$, is divided into enthalpic contributions due to the excluded volume and
electrostatic interactions and into entropic contributions because of small ions, solvent molecules and the polyelectrolyte chain.
Denoting these contributions by $E_{w},E_{e}, S_{i}, S_{s}$, and  $S_{p}$, respectively,
$F^{\star}_{f}$ is given by
\begin{eqnarray}
F^{\star}_{f} - F_{0} &=& E_{w} + E_{e} - T (S_{i} + S_{s} + S_{p}),  \label{eq:free_thermo_exp}
\end{eqnarray}
where $F_{0} = \frac{\rho_{0}}{2} \left( Nw_{pp} + n_{s}w_{ss} \right )$ is the self-energy
contribution arising from the excluded volume interactions characterized by the excluded volume
parameters $w_{ij}$ between species $i$ and $j$. Explicit expressions for different constituents of $F^{\star}_{f}$ are presented in Table ~\ref{tab:t1} in terms of densities and fields at the saddle point. Within the saddle-point approximation, the total free energy ($F_{SCFT}$) of the system is given by $F_{SCFT} \simeq F^{\star}_{a} + F^{\star}_{f}$. In order to compare
the free energies obtained from SCFT and the variational formalism for a given $N$ and $R$,
a single Gaussian chain of contour length $Nl$ in the volume $\Omega$
is chosen as the reference frame, whose free energy is taken to be zero.
This reference free energy of confinement for a single Gaussian
chain has been subtracted from the polymer conformational entropy in Table ~\ref{tab:t1}.
The free energy of confining a single Gaussian chain with $N$ Kuhn segments of length $l$ each in
a spherical cavity of radius $R$ can be computed exactly and is given by\cite{muthuescape}
           \begin{eqnarray}
             F_{gaussian} &=& -\ln \left[\int d\mathbf{r}q_0(\mathbf{r},N)\right] = -\ln \left[ \frac{6\Omega}{\pi^2}\sum_{k=1}^{\infty}\frac{1}{k^2}\exp\left[-\frac{k^2\pi^2 N l^2}{6R^2}\right]\right].
           \end{eqnarray}

\begin{table}
\begin{center}
\begin{tabular*}{1.0\textwidth}{@{\extracolsep{\fill}} | c | c | c | }
  \hline
  Term & SCFT & Variational Formalism  \\
  \hline
    $E_{w}-T S_{s}$ & $\chi_{ps}l^{3}\int d\mathbf{r}  \rho_{p}(\mathbf{r})\rho_{s}(\mathbf{r})  + \rho_{0}\int d\mathbf{r} \eta(\mathbf{r})$  & $\frac{4}{3}\left(\frac{3}{2\pi}\right)^{3/2} (1 - 2\chi_{ps})\sqrt{N}/\tilde{l}_{1}^{3/2}$  \\
    & $ + \int d\mathbf{r}\,\rho_{s}(\mathbf{r})\left \{ \ln\left[\rho_{s}(\mathbf{r})\right]- 1 \right \}$ & $+ \chi_{ps}Nl^{3} - \Omega$\\
  \hline
  $E_{e}$ & $\frac{1}{2} \int d\mathbf{r} \,\psi(\mathbf{r})\rho_{e}(\mathbf{r})$  & $2\sqrt{\frac{6}{\pi}}f^{2}\tilde{l}_{B}N^{3/2}\Theta_{0}(a)/\tilde{l}_{1}^{1/2} $    \\
  \hline
  $- T S_{i}$ &  $\sum_{j=c,+,-}\int d\mathbf{r}\,\rho_{j}(\mathbf{r})\left\{\ln\left[\rho_{j}(\mathbf{r})\right] - 1\right \}$ & $(f N + n_{+})\ln\frac{f N + n_{+}}{\Omega}$ \\
   & & $+ n_{-}\ln\frac{n_{-}}{\Omega} - (f N + n_{+} + n_{-})$\\
  \hline
  $- T S_{p}$ &  $- \ln \left[\int d\mathbf{r}q(\mathbf{r},N)/\int d\mathbf{r}q_0(\mathbf{r},N)\right]$ &  $\frac{3}{2}\left[\tilde{l}_{1} - 1 - \ln \tilde{l}_{1}\right]$ \\
                      $\mbox{}$              &  $ - \rho_{0}\int d\mathbf{r} \eta(\mathbf{r}) - \int d\mathbf{r}\, \left [ \left\{Z_{p}f \psi(\mathbf{r}) + w_{p}(\mathbf{r}) \right \} \rho_{p}(\mathbf{r})\right ]$  & \mbox{}    \\
  \hline
\end{tabular*}
\caption{\label{tab:t1} Comparison of contributions to $F_{f}^{\star}$ in SCFT and variational formalism. }
\end{center}
\end{table}
\vskip0.4cm

\subsection{Variational Formalism}\label{sec:variational}
In variational calculations\cite{muthu04}, a single polyelectrolyte
chain, whose monomers interact with the excluded volume and the electrostatic interactions
in the presence of the small ions is approximated by an \textit{effective} Gaussian chain, whose
conformational statistics are dependent on the different kinds of interactions in the system.
To compute the equilibrium free energy, its variational \textit{ansatz} is minimized with respect to the
variational parameter $l_{1}$, which is related to the radius of gyration ($R_g$) of the chain by $R_g^2 = Nll_1/6$. Physically, this corresponds to the minimization of the free energy of the single chain system with respect to the size of the chain. For the computations of the equilibrium degree of ionization, an additional minimization
of the free energy with respect to the degree of ionization has to be carried out. However, due to the intricate coupling between the size of the chain and the degree of ionization, the minimizations have to be carried out self-consistently.

The variational \textit{ansatz} of the total free energy ($F_{variational}$) is given by $F_{variational} = F^{\star}_{a} + F^{\star}_{f} + \Delta F$, where $\Delta F$ involves one-loop fluctuation corrections,
 addressing the density fluctuations of the small ions, to the free energy.
As mentioned before, the free energy of the adsorbed counterions (i.e., $F^{\star}_{a}$ ) is the same
in both SCFT and the variational formalisms (cf. Eq. (~\ref{eq:free_adsorbed})). The $F_{f}^{\star}$ part of the free energy\cite{muthu04} is
tabulated in Table ~\ref{tab:t1}. The function $\Theta_0(a)$ in Table ~\ref{tab:t1} is a cross-over function given by\cite{muthu87,muthu04}
\begin{eqnarray}
\Theta_0 (a) &=& \frac{\sqrt{\pi}}{2}\left( \frac{2}{a^{5/2}}
- \frac{1}{a^{3/2}} \right) \exp (a) \mbox{erfc} (\sqrt{a}) + \frac{1}{3 a}
+ \frac{2}{a^2} - \frac{\sqrt{\pi}}{a^{5/2}} - \frac{\sqrt{\pi}}{2 a^{3/2}},
\end{eqnarray}
where $a \equiv \kappa^2 N l l_1/ 6$, and $\kappa l$ is the dimensionless inverse Debye length.
Furthermore, $\tilde{l}_{1} = l_{1}/l, \tilde{l}_{B} = l_{B}/l$. The number of salt ions ($n_+, n_-$) are related to the salt concentration ($c_s$) by the relation $Z_+ n_+ = -Z_-n_- = 0.6023c_s \Omega$, where $c_s$ is in units of moles per liter (molarity).  Also, all the terms in the free energies are in units of $k_{B}T$.

In this work, we have ignored one-loop corrections to the free energy within SCFT. However, one-loop corrections to
the free energy coming from the density fluctuations of the small ions, within the variational formalism, is given by
\begin{eqnarray}
\Delta F  &=& -\frac{\Omega \kappa^{3}}{12\pi},  \label{eq:one_loop}
\end{eqnarray}
where $\kappa^{2} = 4\pi l_{B}(fN + n_{+} + n_{-})/\Omega$ and $\kappa$ is the inverse Debye length.

\subsection{Numerical Techniques}\label{sec:numerical}
We solve SCFT Eqs. [(~\ref{eq:saddle_pb}) - (~\ref{eq:field_solv})] within spherical symmetry
(i.e. $\mathbf{r} \rightarrow r = |\mathbf{r}|$), using the Dirichlet boundary conditions for $q(r,t)$ and all the fields except $\eta(r)$. Also, due to the use of spherical symmetry in these calculations, we use
 \begin{eqnarray}
         \frac{\partial \psi(r)}{\partial r}\mid_{r=0} &=& \frac{\partial q(r,t)}{\partial r}\mid_{r=0} = 0    \quad \mbox{for all $t$}.
\end{eqnarray}
Starting from an initial guess for fields,
new fields and densities are computed after solving the modified diffusion and Poisson-Boltzmann equation
 by finite difference methods\cite{recipebook}. Broyden's method\cite{recipebook} has been used to solve the set of non-linear equations. The equilibrium value of the degree of ionization ($f^\star$) is obtained
after minimizing the free energy with respect to $f$. We carry out the
numerical minimization of free energy over $f$ using Brent's method\cite{recipebook}. The results presented in this paper were obtained by using a grid spacing of $\Delta r = 0.1$ and contour steps of $\Delta t = 0.01$.

On the other hand, the self-consistent minimization of the free energy in the variational
method has been carried out by assuming a uniform expansion of the chain within spherical symmetry. In this formalism, the free energy is minimized simultaneously with
respect to $f$ and $l_{1}$, and both these quantities at equilibrium ($f^\star, l_{1}^\star$) are computed self-consistently. The radius of gyration of the chain (which is confined to a finite volume, $\Omega = 4\pi R^{3}/3$) is obtained from the equilibrium value of the expansion factor  $l_{1}^\star$. For these calculations, the upper bound for the radius of gyration of the chain is specified to be the radius of the confining volume (i.e, $R_g \le R$) to mimic the confinement effects. Also, the Kuhn step length $l$ is taken to be unity in both variational as well as SCFT calculations.

\section{Results}
\setcounter {equation} {0} \label{sec:results}
\subsection{Degree of ionization}

We have carried out an exhaustive comparison between the SCFT and variational formalisms
by calculating the equilibrium degree of ionization ($f^\star$) of a negatively charged
single flexible polyelectrolyte chain (i.e. $Z_{p} = -1, Z_{c} = 1$)
in the presence of a monovalent salt. The equilibrium degree of ionization is determined as a function of the strength
of the electrostatic interaction (or Coulomb
strength) which is proportional to the Bjerrum length $l_B$ for a given solvent. For both cases,
the effective charge expectedly decreases
(Fig. ~\ref{fig:lb_effect_k3}) with higher Coulomb strengths that help a progressively larger
degree of adsorption of counterions on the chain backbone.
However, $f^\star$ obtained from the variational procedure is systematically higher than that from 
SCFT. It is to be noted that the degree of ionization is essentially zero in SCFT for experimentally relevant values of $l_B/l$ (around 3 for aqueous solutions), whereas $f^\star$ is reasonable in the variational theory. 
Although both theories use different approximations and computational procedures, there is one major conceptual input that distinguishes these theories. Whereas the variational formalism of Ref. ~\cite{muthu04} includes the density fluctuations of the small ions as one-loop corrections to the free energy, the SCFT does not address these fluctuations. In an effort to quantify  the consequences of small ion density fluctuations and then compare the consequences of the rest of the terms in the variational theory against SCFT (which does not contain small ion density fluctuations by construction), we subtract $\Delta F$ from $F_{variational}$ and then compute $f^*$. 
The results are given in Figs. ~\ref{fig:lb_effect} and ~\ref{fig:compare_free_all}. 

Remarkably, in different conditions corresponding to widely varying degrees of
confinement, the $f^\star$
obtained by the minimization of SCFT free energies is indistinguishable from that
obtained using variational free energies without one-loop corrections (i.e., $F_{variational}-\Delta F$).
We demonstrate this in Fig. ~\ref{fig:lb_effect} where we have plotted
$f^{\star}$ as a function of $l_{B}/l$ for different spherical volumes (i.e., different $R$).
Thus we arrive at two conclusions: (a) density fluctuations of small ions included in the full variational formalism
contribute significantly in determining the equilibrium degree of ionization and lead to better values of $f^\star$
than SCFT, and (b) the value of $f^\star$ is remarkably indistinguishable between the SCFT and the variational formalism with
deliberate suppression of small ion density fluctuations. The first conclusion can be readily rationalized as follows by
considering the two curves in Fig. ~\ref{fig:lb_effect_k3}.

The increase in $f^\star$ with the inclusion of $\Delta F$ can be
understood by the fact that the density fluctuations of the small ions lower the free energy, and its contribution to the total
free energy increases with the increase in the number of ``free'' ions (goes like $-n^{3/2}$ in salt-free case,
where $n$ is the number of ``free'' ions - cf. Eq. (~\ref{eq:one_loop})). For higher values of $l_B/l$ (above $4$),
all the counterions are adsorbed on the chain so that the degree of ionization of the chain is zero irrespective of the
density fluctuations. In contrast, for lower values of $l_B/l$ (below $0.5$), the chain is fully ionized and the effect of
the density fluctuations of the small ions on the effective degree of ionization is minimal. However, for the intermediate
values of $l_B/l$, the density fluctuations of the small ions affect the degree of ionization significantly and non-monotonic
deviations from the SCFT results as a function of $l_B/l$ are observed in this regime. Also, term-by-term comparison of the free energy components reveals that
the discrepancy arises solely
due to the term accounting for the density fluctuations of the small ions.
This disagreement highlights the fact that the effect of density fluctuations of the small ions
is not included in SCFT within the saddle point approximation.

The above second conclusion requires further scrutiny. The remarkable agreement between the two formalisms is surprising, since these
theories use different approximations and different computational procedures. In the variational formalism of
Ref. \cite{muthu04}, that is used in the present paper, the chain swelling due to electrostatic interaction is assumed to be
spherically uniform at all length scales and at the level of Debye-H\"{u}ckel potential between the segments. However, this scheme is
more tractable analytically with different contributing factors (Table ~\ref{tab:t1}) having explicit physical interpretation.
On the other hand, in SCFT, the electrostatic interaction is at the nonlinear Poisson-Boltzmann level, and the chain expansion is
addressed at all local length scales through fields generated by intersegment potentials. Although the chain conformations are not
readily accessible in the standard version of SCFT used here, the free energy of the system can be calculated 
and its resolution into entropic and
enthalpic parts is possible. In view of such apparently divergent approaches in SCFT and the variational formalisms, we now
proceed to make quantitative comparisons between the two in terms of the various contributing factors.

\subsection{Term-by-term comparison of free energy: SCFT and variational formalism}

To assess the approximations used in the variational theory and to find out the origin of the
remarkable agreement in terms of the equilibrium effective charge ($f^\star$)
obtained from SCFT and variational theory
(with deliberate suppression of one-loop corrections for small ion density fluctuations),
we have compared individual contributions to
the free energies in these two formalisms. Before presenting the numerical results, the role of
different contributions in driving the counterion adsorption can be understood qualitatively by considering the following physical picture.

The driving forces for the counterion adsorption are the formation of ion-pairs due to the presence of strong attractive interactions in the process (self-energy of the dipoles) and the decrease in intramolecular electrostatic repulsions (compared with a fully
ionized chain, where these repulsions are the strongest). However, an extensive counterion
adsorption on the chain backbone is unfavorable due to the loss in translational degrees of freedom
of the ``free'' counterions. Another factor, which plays a role in this competition of the energy and
entropy is the translational entropy of the ``adsorbed'' counterions.
This entropic feature alone favors a state where half of the charges on the chain backbone are free and the other half
adsorbing the counterions. With an
increase in the electrostatic interaction strength (i.e., Bjerrum length), the driving forces for the
counterion adsorption increase and drive more and more counterions to the backbone.
On the other hand, an extensive counterion adsorption leads to the chain contraction due to lower intramolecular
electrostatic repulsions (a result of the lower number of bare charged sites) even when the electrostatic interaction
strength is high. However, we show that the
counterion adsorption leading to the lowering of the effective charge (that decreases the electrostatic energy due to the
formation of ion-pairs) has a
bigger effect than the chain contraction (which affects the chain conformational entropy and polymer-solvent entropy) or
the increase of intramolecular electrostatic repulsions among the unadsorbed segments as we
gradually increase the Bjerrum length. Of course, in addition, correlations of small ion density fluctuations also
contribute to $f^\star$, in the full variational calculation. Numerical results on the relative importance of the various
contributions to the
total free energy along with their role in driving the counterion adsorption are presented below.

To start with, in Fig. ~\ref{fig:compare_free_all}, we have plotted the total free
energy calculated in both methods for the following set of parameters:
$Z_{p} = -Z_{c} = -1, R/l = 10, N = 100, c_s = 0.1M, \delta = 3$ and $\chi_{ps} = 0.45$. It is clear that the
total free energies obtained from SCFT and the variational theory are in quantitative agreement with each other.

To analyze this striking agreement between the two methods we focus on the individual components of the free
energy as tabulated in Table ~\ref{tab:t1}. In Figs. ~\ref{fig:compare_free_major}
and ~\ref{fig:compare_free_minor} we have compared these different
constituents of the free energy obtained from both SCFT and the variational formalisms for low monomer
densities. It is evident that both theories predict that the major contributions to the free energy
are from the ion-pair energy(Fig. ~\ref{fig:compare_free_major}a), the ``adsorbed'' counterion translational
entropy (Fig. ~\ref{fig:compare_free_major}b), the polymer-solvent interaction energy and the solvent
entropy (Fig. ~\ref{fig:compare_free_major}c); and the ``free'' ions translational entropy (Fig. ~\ref{fig:compare_free_major}d).
Contributions due to the chain conformational entropy (Fig. ~\ref{fig:compare_free_minor}a) and the electrostatic
energy (Fig. ~\ref{fig:compare_free_minor}b) are almost negligible (less than $0.1\%$ in the total free energy)
as compared to others.
For low monomer densities (monomer volume fractions lower than $0.1$), the dominant contributions to the total free energies come from
the polymer-solvent interaction energy and the solvent entropy. For the particular single chain
dilute system investigated here, these contributions
account for more than $50 \%$ of the total free energy. Although large, these contributions
are found to be almost insensitive to $f$. For example, polymer-solvent interaction energy changes only by less than
$0.5 k_B T$ when $l_B/l$ is varied from $0.2$ to $5.0$. On the other hand, $f^\star$ changes from $1$ to $0$ in the same range of
$l_B/l$. In fact, the $f$ dependent terms which contribute significantly to the total
free energy are the ion-pair energy and the ``free'' ions translational entropy. At lower electrostatic
interaction strengths (i.e., low $l_B/l$), the translational entropy of the ``free'' ions dominates
and at higher electrostatic strengths, the ion-pair energy term contributes significantly to the free energy.
Together, these two contributions account for as high as $99 \%$ of the $f$ dependent part in the
total free energy (cf. Figs. ~\ref{fig:compare_free_all}, ~\ref{fig:compare_free_major}a and ~\ref{fig:compare_free_major}d).
Relatively very small contributions ($\sim 1 \%$) to the free energies come from the translational entropy
of the ``adsorbed'' ions. We will see below, however, that the relative importance of a particular contribution
in determining the equilibrium degree of ionization is not necessarily related to its actual contribution to the
total free energy.

We now discuss the various trends seen in Figs. ~\ref{fig:compare_free_major} and ~\ref{fig:compare_free_minor},
based on conceptual arguments aided by the different terms in
Table ~\ref{tab:t1}. Intuitively, stronger ion-pair energy should promote counterion adsorption. As $l_B/l$ is increased, the
energy due to counterion adsorption should decrease monotonically, as seen in Fig. ~\ref{fig:compare_free_major}a.
From Eq. ~\ref{eq:ea}, it is clear that the ion-pair energy (note the negative contribution) favors counterion adsorption
with a linear dependence on $f$ and the Coulomb strength $l_B$, hence a progressive gain in adsorption energy
with increasing Coulomb strength (Fig. ~\ref{fig:compare_free_major}a). On the other hand, the counterion adsorption
is opposed by the translational entropy of the ``free'' ions (see the expression for $TS_i$ in Table ~\ref{tab:t1}), and
hence a progressive loss of the part of the free energy related to the translational entropy
(Fig. ~\ref{fig:compare_free_major}d). The plateau in Fig. ~\ref{fig:compare_free_major}d arises due to
the completion of the adsorption of all counterions which limits the loss in the number of free counterions
(which is the number of salt ions) with increasing Coulomb strength. However, there is no plateau in Fig. ~\ref{fig:compare_free_major}a,
because even after all counterions are adsorbed, the ion-pair energy continually decreases due to an increasing Coulomb strength.
In addition, the translational entropy of the ``adsorbed'' counterions ($-TS_a$) drives the adsorption toward $f^\star = 0.5$ to optimize
this part of the entropy (cf. Eq. ~\ref{eq:sa}). Physically, it can be understood from the fact that the complete adsorption of the
counterions leads to the lowering of the translational entropy of the ``adsorbed'' counterions due to the unavailability
of sites. Similarly, a complete desorption of the counterions also leads to the lowering of translational entropy of the ``adsorbed'' counterions due to the unavailability of the ``adsorbed'' counterions on the chain backbone. For a given $N$, the
translational entropy of the ``adsorbed'' counterions is optimum at $f^\star = 0.5$. We note, however, that the other two contributions
might overwhelm $-TS_a$ so that
at equilibrium it is not necessarily at its minimum (Fig. ~\ref{fig:compare_free_major} b). With varying Coulomb strength, $-TS_a$ is minimum at around
$l_B/l=0.8$ at which $f \simeq 0.5$, which is prevalently determined by the first two components mentioned above.
Role of other contributions, i.e., the polymer-solvent interaction energies and the solvent entropy (i.e., $E_w-TS_s$),
the electrostatic energy involving  the ``free'' ions and the monomers ($E_e$) and the conformational entropy of the chain ($-TS_p$) is
miniscule in driving the counterion adsorption in a particular direction. However, these three contributions dictate
the effective size of the chain (through $l_1$) at the equilibrium (note the dependence of these terms on $l_1$ in Table ~\ref{tab:t1}).
Further, the equilibrium counterion distribution specified by the first three contributions stipulates the actual contributions of the
last three parts of the free energy at equilibrium. We have noticed before that with increasing Coulomb strength ($l_B/l$) the number of free
counterions (and, therefore, the effective charge of the chain) decreases. Due to a decreasing electrostatic repulsion between the
monomers, the polymer chain contracts progressively until it reaches its Gaussian size at zero effective charge.
Consequently, there is less mixing between the polymer and the solvent at higher Coulomb strengths leading to a gradual
loss of polymer-solvent interaction energy (Fig.  ~\ref{fig:compare_free_major}c) which saturates to a plateau when all counterions adsorb, the physical
condition that creates plateaus in all these curves. Also accompanying the decreasing size of the chain there is a gain
in conformational entropy (which is maximum at the Gaussian size) observed in Fig. ~\ref{fig:compare_free_minor}a.
In addition, a gradual decrease of the effective
charge of the chain progressively reduces the electrostatic energy penalty observed in Fig. ~\ref{fig:compare_free_minor}b.
However, this effect is very small compared to the lowering of electrostatic energy due to the formation of ion-pairs as mentioned
earlier.

The quantitative agreement between the first three contributions to the free energy in two formalisms explains the observed agreement in
the results obtained for $f^\star$ (Fig. ~\ref{fig:lb_effect}).
Despite $E_e$ having a negligible contribution to the total free energy (less than $0.1 \%$),
the comparison reveals that the Debye-H\"{u}ckel estimate for the electrostatic energy ($E_{e}$) used in the
variational formalism is an overestimation (as large as five times the full, non-linear Poisson-Boltzmann at low $l_B/l$).
In other words, the Debye-H\"{u}ckel approximation underestimates the degree of screening, which is in agreement with other
theoretical\cite{stigter95} and simulation results\cite{kremer96}. Note that the
electrostatic energy in Fig. ~\ref{fig:compare_free_minor}b includes all the charged species in the system except the
ion-pairs formed on the chain by the adsorbing counterions. Nevertheless, contributions due to the electrostatic energy to
total free energy are almost negligible and hence, do not affect $f^{\star}$ significantly.

We have also carried out the same comparison between the two formalisms at higher monomer
densities (above monomer volume fractions of $0.1$).
It is found that the discrepancy
in the polymer-solvent interaction energy and the solvent entropy between the two schemes is significant 
(see Fig. ~\ref{fig:compare_ew_tss_high_density}). 
All other contributing factors are essentially the same between the two theories. 
The origin of this discrepancy lies in the expansion of the $(1-\rho_p)\log (1-\rho_p)$ term, which is carried up to only terms quadratic
in polymer density in the variational calculations (Appendix-B). The higher order terms in the expansion are ignored in the variational
calculations to carry out the analysis analytically, which limits the
applicability of the variational theory to sufficiently low monomer concentrations.
The discrepancy clearly highlights the breakdown of the variational procedure at high densities and questions the
use of an effective excluded volume parameter in variational calculations.
However, we have not made an attempt to compute the boundary of the disagreement between the theories
because our main focus of this article is on $f^{\star}$ which is insensitive to this discrepancy.
Also, the variational formalism predicts the
polymer-solvent interaction energy and the solvent entropy to be completely independent of the electrostatic
interaction strength $l_B/l$ in the high density regime, in contrast to SCFT predictions of a weak dependence
on $l_B/l$ (see Fig. ~\ref{fig:compare_ew_tss_high_density}).  This is a result of the constraint $R_g \le R$ used
in variational calculations for mimicking the confinement effects and shows the inability of the constraint
to capture the confinement effects in an appropriate fashion. While the radius of gyration
$R_g$ of the chain follows readily from $l_1$ in the variational formalism, it is nontrivial to compute this quantity in
SCFT. In view of this, we have not addressed $R_g$ in the present paper.

Finally, we remark on the experimental relevance of the radius parameter $R$ for the confining cavity. The variational calculation readily gives $R_g$ without any confinement for fixed values of monomer density and other parameters such as $l_B, \chi_{ps},$ and $c_s$. Knowing this result, we have investigated the role played by the cavity radius $R$ in the above analysis. If $R$ is larger than $R_g$, then the above conclusions are relevant to unconfined dilute polyelectrolyte solutions. On the other hand, if $R$ is less than $R_g$, then confinement effect is manifest and now our results are relevant to a polyelectrolyte chain inside a spherical pore. As an example, for $N=100, c_s = 0.1M, \chi_{ps} = 0.45$, and $\delta = 3$, the calculated value of $R_g/l$ from the variational procedure depends on $l_B$ and attains a maximum value of $7.29$, whereas $R/l= 10.0$ in Figs. ~\ref{fig:compare_free_all}-~\ref{fig:compare_free_minor}. Therefore, the conclusions drawn above based on these figures are generally valid for dilute polyelectrolyte solutions. On the other hand, $R/l = 4.0$ in 
Fig. ~\ref{fig:compare_ew_tss_high_density}, whereas the maximum value of $R_g$ would have been $5.92$ if confinement were to be absent. Under these conditions  the conclusions regarding the discrepancy between SCFT and variational theory is pertinent to a polyelectrolyte chain confined inside a spherical cavity. 

\section{Conclusions}
\setcounter {equation} {0} \label{sec:conclusions}
In summary, we have computed the effective charge of a single flexible polyelectrolyte chain
using SCFT and compared it with the results obtained from a variational theory. It is found that for all sets of
parameters, the effective degree of ionization ($f^\star$) computed from SCFT and the variational theory
is in quantitative agreement if one-loop fluctuation corrections are deliberately suppressed in the latter.
The origin of this agreement lies in the fact that $f^\star$ is determined as an interplay of the ion-pair energy
and the translational entropy of the ``adsorbed'' counterions as well as of all ``free'' ions.
The conformational entropy of the chain, the electrostatic energy involving the ``free'' ions and the chain, the polymer-solvent interaction energy and the solvent entropy do not play significant roles
in affecting $f^\star$.

The comparison of different components in free energy reveals that the Debye-H\"{u}ckel
approximation underestimates screening effects as compared to the Poisson-Boltzmann theory. Despite the fact that there are
small discrepancies in the different contributing factors to the total free energy, the effective degree of ionization ($f^\star$)
comes out to be the same in SCFT and the truncated variational theory. 
Furthermore, the density fluctuations of the ``free'' ions, which are included in the full variational theory, are predicted to
increase the equilibrium degree of ionization.
As this latter effect is not captured by SCFT calculations within the saddle-point approximation and due to the close
agreement between SCFT and the variational theory for all other contributing factors, the variational
theory appears to be a very useful tool for a quick, easy and transparent estimation of $f^\star$.

\section*{ACKNOWLEDGEMENT}
\setcounter {equation} {0} \label{acknowledgement}
 We acknowledge
financial support from the National Institute of Health (Grant No. R01HG002776),
National Science Foundation (Grant No. DMR-0605833), AFOSR (FA9550-07-1-0347), 
and the Material Research Science and Engineering Center at the University of
Massachusetts, Amherst.

\renewcommand{\theequation}{A-\arabic{equation}}
  \setcounter{equation}{0}  
  \section*{APPENDIX A : SCFT for ``permuted'' charge distribution } \label{app:A}
Here, we present a brief description of SCFT for a single flexible
polyelectrolyte chain having a fixed degree of ionization ($=f$) in the presence of salt ions and solvent molecules. We start from
the Edward's Hamiltonian, written as
\begin{eqnarray}
       \mbox{exp}\left(-\frac{F - F_{0}} {k_{B}T}\right )& = & \frac {\mbox{exp}\left[-E_{a}/k_{B}T \right]}{\mu \prod_{j}n_{j}!}\int D[\mathbf{R}] \int \prod_{j} \prod_{m=1}^{n_{j}} d\mathbf{r}_{m} \quad \left[\mbox{exp} \left \{-\frac {3}{2 l}\int_{0}^{Nl}
        dt\left(\frac{\partial \mathbf{R}(t)}{\partial t} \right )^{2} \right . \right . \nonumber \\
&& - \chi_{ps}b^{3}\int d\mathbf{r} \hat{\rho}_{p}(\mathbf{r})\hat{\rho}_{s}(\mathbf{r})
 \left . - \frac{1}{2}\int d\mathbf{r}\int d\mathbf{r}\,'\frac{\hat{\rho}_{e}(\mathbf{r})\hat{\rho}_{e}(\mathbf{r}\,')}{4\pi \epsilon_0 \epsilon k_{B}T |\mathbf{r}-\mathbf{r}\,'|}  \right \}\nonumber \\
&& \left . \prod_{\mathbf{r}}\delta\left(\hat{\rho}_{p}(\mathbf{r}) + \hat{\rho}_{s}(\mathbf{r}) - \rho_{0}
\right) \delta\left(\frac{1}{l}\int_{0}^{Nl} dt \, y(t) - fNe \right)\right]_{y}, \label{eq:parti_sing}
\end{eqnarray}
where $\mathbf{R}(t)$ represents the position vector for the $t^{th}$ segment and subscripts $j = s,c,+,-$.
In Eq. (~\ref{eq:parti_sing}) $k_{B}T$ is the Boltzmann constant times absolute
temperature. In writing the interaction energies between the polyelectrolyte segments and small ions,
we have taken the small ions to be point charges so that
they have zero excluded volume, and hence, interactions are purely electrostatic in nature. As we consider the ``permuted''
charge distribution, the partition function has an additional sum over all possible locations of the ``adsorbed''
ions on the backbone, which appears as an average over the parameter $y$ in Eq. (~\ref{eq:parti_sing}).
We define the average over $y$ as $[\cdots]_{y} = \int dy [\cdots] g(y)$, where $g(y) = f\delta(y(t)-1) + (1-f)\delta(y(t))$.

In the above equation, the microscopic densities are defined as
        \begin{eqnarray}
    \hat{\rho}_{p}(\mathbf{r})  &=& \frac{1}{l}\int_{0}^{Nl} dt \, \delta (\mathbf{r}-\mathbf{R}(t)), \\
     \hat{\rho}_{j}(\mathbf{r}) &=&  \sum_{i=1}^{n_{j}} \delta (\mathbf{r}-\mathbf{r}_{i}) \quad \mbox{for} \quad j = s,c,+,-,\\
\hat{\rho}_{e}(\mathbf{r}) &=& e \left [ \frac{1}{l}\int_{0}^{Nl} dt \, Z_{p} y(t) \delta (\mathbf{r}-\mathbf{R}(t)) + \sum_{j = c,+,-}Z_{j}\hat{\rho}_{j}(\mathbf{r}) \right ],
\end{eqnarray}
where $\hat{\rho}_{p}(\mathbf{r}), \hat{\rho}_{j}(\mathbf{r})$ and $\hat{\rho}_{e}(\mathbf{r})$
stand for the monomers, small molecules (both ions and solvent molecules) and the local charge density, respectively.
The Dirac delta functions involving microscopic densities
enforce the incompressibility condition at all points in the system ($\rho_{0}$
being the total number density of the system). The delta function involving $y$ is a constraint that for all the
charge distributions to be considered for one particular value of $f$, the net charge on the chain must be a constant ($= fNe$).
Taking different charge distributions of the chain for the same net charge ($=fNe$) and a particular chain conformation to be degenerate,
the partition function is divided by the number of ways ($\mu$) the ``adsorbed'' counterions can be distributed along the chain.
If $M$ out of total $N$ sites on the backbone are occupied at any particular instance, then $\mu$ is given by $\mu = N!/(M! (N-M)!)$
so that $1- f = M/N$.

The dimensionless Flory parameter for chemical mismatch, $\chi_{ps}$, is given by $\chi_{ps}l^{3} = w_{ps} - (w_{pp} + w_{ss})/2$,
where $w_{pp}, w_{ss}$, and $w_{ps}$ are the excluded volume parameters, which characterize
the short range excluded volume interactions of type monomer-monomer, solvent-solvent and monomer-solvent,
respectively. $F_{0}$ and $E_{a}$ are the self-energy and ion-pair energy contributions, respectively, given by

\begin{eqnarray}
       \frac{F_{0}} {k_{B}T} & = & Nw_{pp} + n_{s}w_{ss}, \label{eq:parti_den}\\
       \frac{E_{a}}{k_{B}T} &=& -(1-f)N\delta l_{B}/l,
\end{eqnarray}
where $\delta = \epsilon l /\epsilon_{l} d$, $\epsilon_{l}$ and $d$ being the local dielectric constant and the
dipole length, respectively, is used to characterize the formation of an ion-pair on the bakbone due to ``adsorbed'' counterion.

 Now, using the methods of collective variables and the Hubbard-Stratonovich transformation\cite{fredbook,rajeev08} for the
electrostatic part in Eq. (~\ref{eq:parti_sing}), the partition function can be written as integrals over the collective
densities and corresponding fields so that Eq. (~\ref{eq:parti_sing}) becomes

  \begin{eqnarray}
       \exp\left(-\frac{F - F_{0}} {k_{B}T}\right )& = & \int D[w_{p}] D[\rho_{p}]D[\psi] D[\eta] du D[w_{s}] D[\rho_{s}] \, \left[\exp \left \{-\frac{H_{scf}}{k_{B}T}\right \}\right]_{y}. \label{eq:parti_scf}
\end{eqnarray}
Here, $w_{p},w_{s}$ are the fields experienced by the monomers and solvent, respectively, and $\rho_{p},\rho_{s}$
represent their respective collective densities. All charged species (excluding the ion-pairs formed due to adsorption of counterions)
experience a field $\psi$ (which is equivalent to the electrostatic potential). $\eta$ and $u$ are Lagrange's multipliers
corresponding to, respectively, the incompressibility and net charge constraints in the partition function. We must stress here
that this procedure is equivalent to introducing collective fields and densities for small ions instead of using the Hubbard-Stratonovich
transformation for the electrostatics part.

  Within SCFT, the functional integrals over the fields and the densities are approximated by the value of the integrand
  at the extremum (also known as the saddle-point approximation). Extremizing the integrand leads to a number of non-linear
equations for the fields and the densities. The saddle point approximation with respect to $u$ gives equations similar to a ``smeared'' charge
distribution, where every monomer has a charge ($=f e$). The extremization with respect to $\psi,w_p,\rho_p,\eta,w_s$, and $\rho_s$,
leads to the saddle point specified by Eqs. (~\ref{eq:saddle_pb} - ~\ref{eq:den_eq}, ~\ref{eq:saddle_exp1} - ~\ref{eq:field_solv}).
Using these saddle point equations and employing the Stirling's approximation for $\ln n!$, we obtain the approximated free energy, i.e.,
$F - F_{0} \simeq H_{scf}^{\star}$, as presented in section ~\ref{sec:scft} after taking $k_B T = 1$.  The superscript $\star$ represents
the saddle point estimate of the free energy.

\renewcommand{\theequation}{B-\arabic{equation}}
  \setcounter{equation}{0}  
  \section*{APPENDIX B : Variational theory }

In this appendix, we present the procedure to obtain the variational free energy as presented in Ref. \cite{muthu04} in the absence of
ion-pair correlations. In Ref.\cite{muthu04}, it has been assumed that the counterions from the polyelectrolyte are indistinguishable from
the counterions from the salt. So, we start from a partition function similar to
Eq. (~\ref{eq:parti_sing}) with the solvent, counterions (from the polyelectrolyte and the salt), coions and the chain
as distinguishable species. After using collective variables, the partition function can be written as
\begin{eqnarray}
       \mbox{exp}\left(-\frac{F - F_{0}} {k_{B}T}\right ) & = & \int D[w_{p}] D[\rho_{p}]D[\eta] \prod_{j}D[\rho_{j}] D[w_{j}] \, \mbox{exp} \left \{-\frac{h}{k_{B}T}\right \}, \label{eq:parti_vari}
\end{eqnarray}
where $j=s,c,-$ and, where the integral over $u$ has already been evaluated by the saddle point method so that the functional $h$
corresponds to a single chain with a ``smeared'' charge distribution. Here, we have introduced collective fields and densities for small
ions instead of using the Hubbard-Stratonovich transformation for the electrostatic part (as already pointed out in Appendix-A). This is
the analog of Eq. (~\ref{eq:parti_scf}) in SCFT. Now, evaluating the path integrals over $w_{j}$ by the saddle point method,
Eq. (~\ref{eq:parti_vari}) can be written in terms of the densities $\rho_{j}$. Functional integrals over $\eta$ and $\rho_s$ can be carried out
in a trivial way. To carry out functional integrals over small ion densities, i.e, the $\rho_j \log \rho_j$ terms, which emerge after
integrations over fields $w_j$, are expanded up to the quadratic terms after writing
$\rho_{j}(\mathbf{r}) = n_{j}/\Omega + \delta \rho_{j}(\mathbf{r}) $ so that $\int d\mathbf{r} \delta \rho_{j}(\mathbf{r}) = 0$, and
the resulting integrals are Gaussian. This procedure also gives one-loop corrections to the free energy coming from the small ions density
fluctuations ($\Delta F/k_{B}T$). Now, expanding the $(1- \rho_p) \log (1-\rho_p)$ term up to the quadratic terms in $\rho_p$, the problem
of carrying out the functional integrals over $w_p$ and $\rho_p$ is equivalent to a single chain problem whose monomers interact
with each other via a renormalized excluded volume parameter and an electrostatic potential. The renormalized excluded volume parameter
comes out to be $w = 1 - 2\chi_{ps}$ and the electrostatic potential comes out to be the Debye-H\"{u}ckel potential, where the inverse
Debye length ($\kappa$) depends on the ``free'' ions only. Eventually, Eq. (~\ref{eq:parti_vari}) becomes
\begin{eqnarray}
       \mbox{exp}\left(-\frac{F - F_{0}-\delta F} {k_{B}T}\right ) & = & \frac{1}{\mu}\mbox{exp} \left \{-\frac{E_a - TS_i}{k_{B}T}\right \} \int D[w_{p}] D[\rho_{p}] \, \mbox{exp} \left \{-\frac{H_{var}}{k_{B}T}\right \}, \label{eq:parti_vari_sing}
\end{eqnarray}
where $-TS_i$ is the translational entropy of the ``free'' ions as presented in Table ~\ref{tab:t1} for the variational theory.
Now, writing the Hamiltonian of a single polyelectrolyte chain using an effective excluded parameter ($w$) and the Debye-H\"{u}ckel
potential\cite{muthu04}, the functional integrals over $w_p$ and $\rho_p$ can be computed using the variational technique
developed by Muthukumar\cite{muthu87}. Taking $k_{B}T = 1$, the results are presented in section ~\ref{sec:variational}.

\section*{REFERENCES}
\setcounter {equation} {0}
\pagestyle{empty} \label{REFERENCES}

\newpage
\section*{FIGURE CAPTIONS}
\pagestyle{empty}

\begin{description}
\item[Fig. 1.:] Comparison of SCFT and the variational formalism (with one-loop corrections) to illustrate the effect of
correlations among small ions on the effective degree of ionization ($f^\star$). $Z_p = -Z_c = -1, R/l = 10, N = 100, c_{s} = 0.1M, \chi_{ps} = 0.45$ and $\delta = 3$.
\end{description}

\begin{figure}[h]
 \begin{center}
     \vspace*{1.0cm}
      \begin{minipage}[c]{15cm}
     \includegraphics[width=15cm]{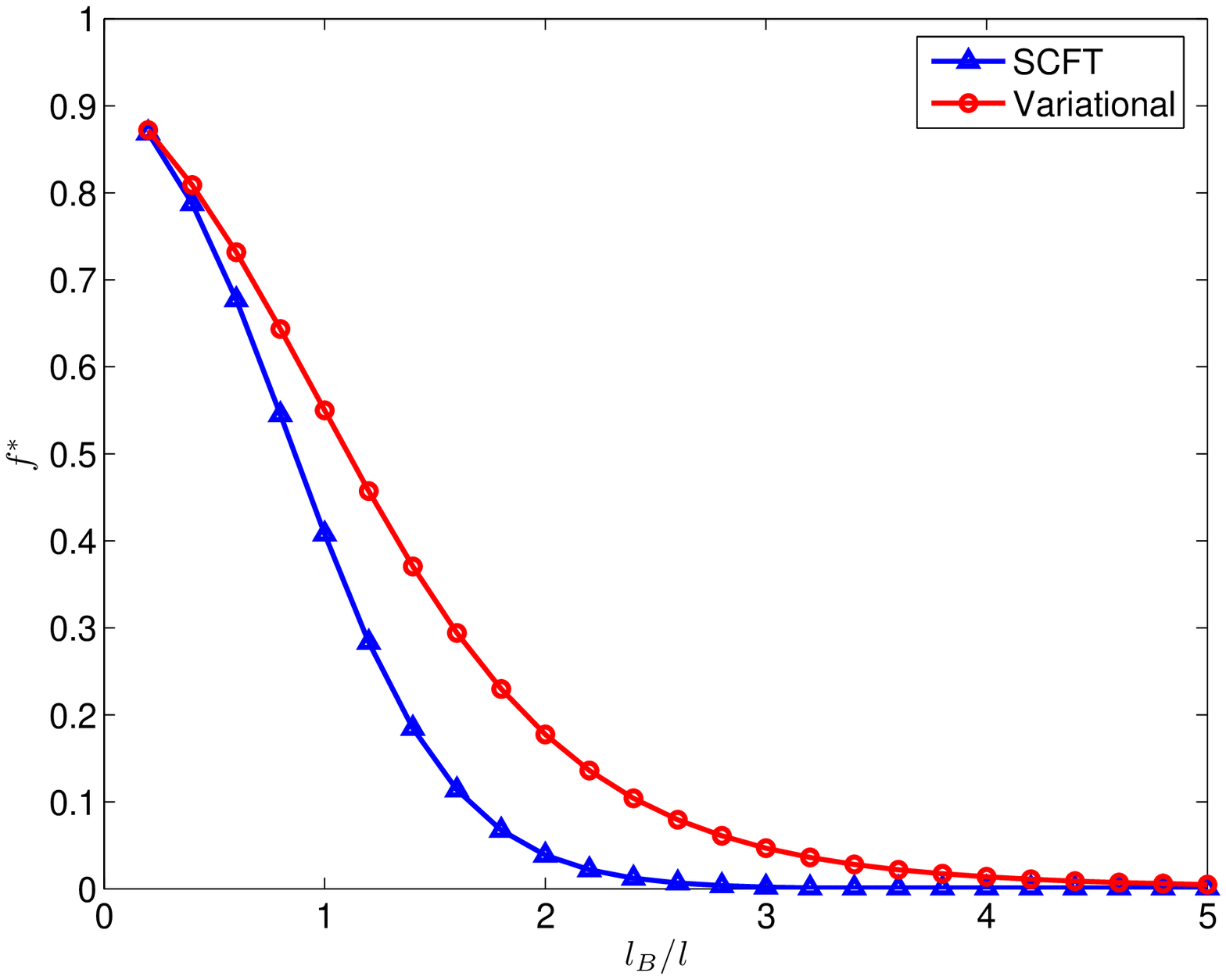}
    \end{minipage}
\caption{} \label{fig:lb_effect_k3}

\end{center}
\end{figure}

\begin{figure}[h]
 \begin{center}
     \vspace*{1.0cm}
      \begin{minipage}[c]{15cm}
     \includegraphics[width=15cm]{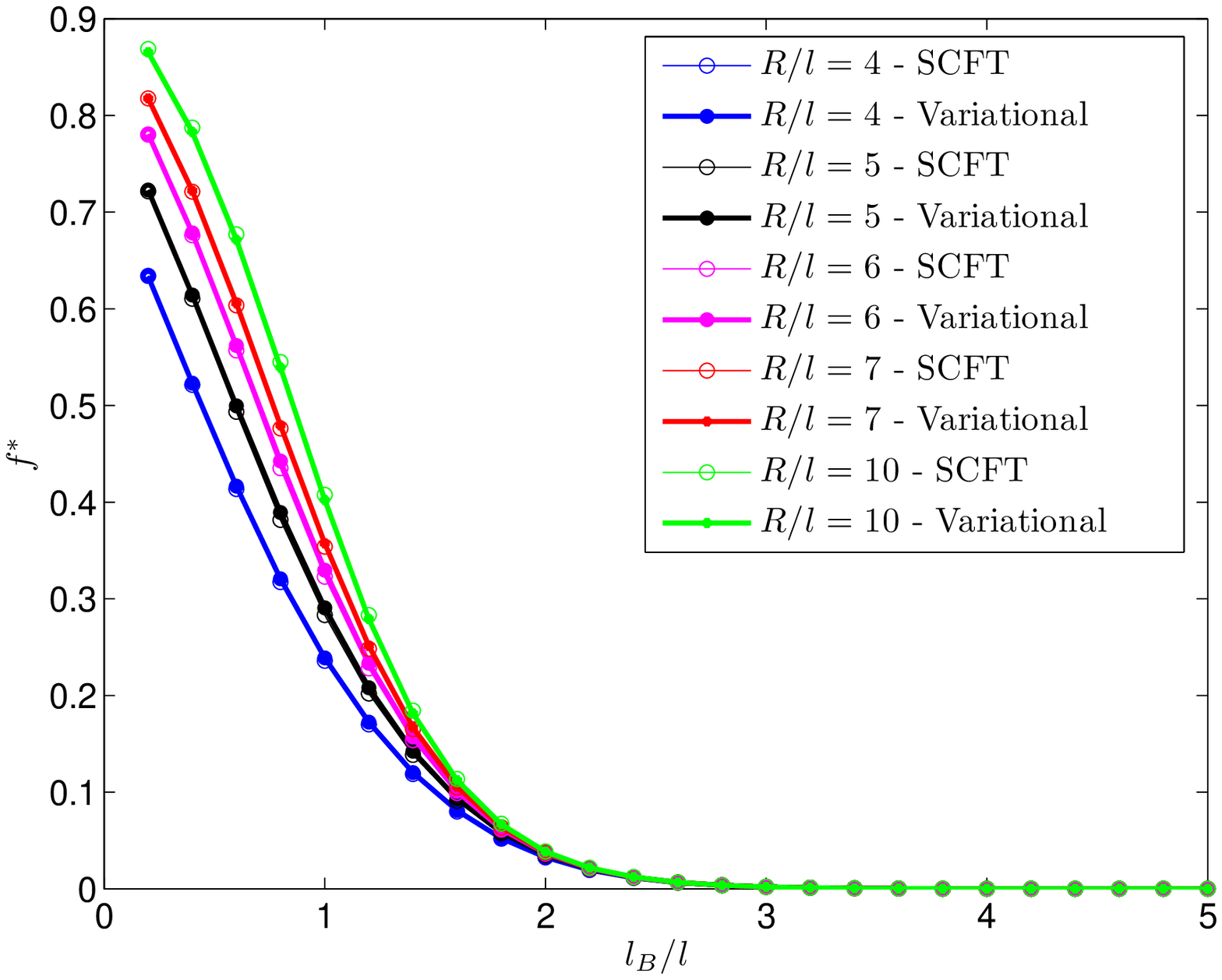}
    \end{minipage}
\caption{} \label{fig:lb_effect}

\end{center}
\end{figure}

\begin{description}
\item[Fig. 2.:] Comparison of $f^\star$ computed using SCFT and the variational formalism (without one-loop correction)
for different values of $R$ and $l_{B}/l$. $Z_p = -Z_c = -1, N = 100, c_{s} = 0.1M, \chi_{ps} = 0.45$ and $\delta = 3$.  Plot for
SCFT when $R/l = 10$ is the same as in Fig. ~\ref{fig:lb_effect_k3}.
\end{description}

\begin{description}
\item[Fig. 3.:] Comparison of total free energies (at equilibrium, for $f=f^\star$) obtained from SCFT and the
variational calculations (without one-loop correction). $Z_p = -Z_c = -1, R/l = 10, N = 100, c_{s} = 0.1M, \chi_{ps} = 0.45$ and $\delta = 3$.
\end{description}

\begin{figure}[h]
 \begin{center}
     \vspace*{1.0cm}
      \begin{minipage}[c]{15cm}
     \includegraphics[width=15cm]{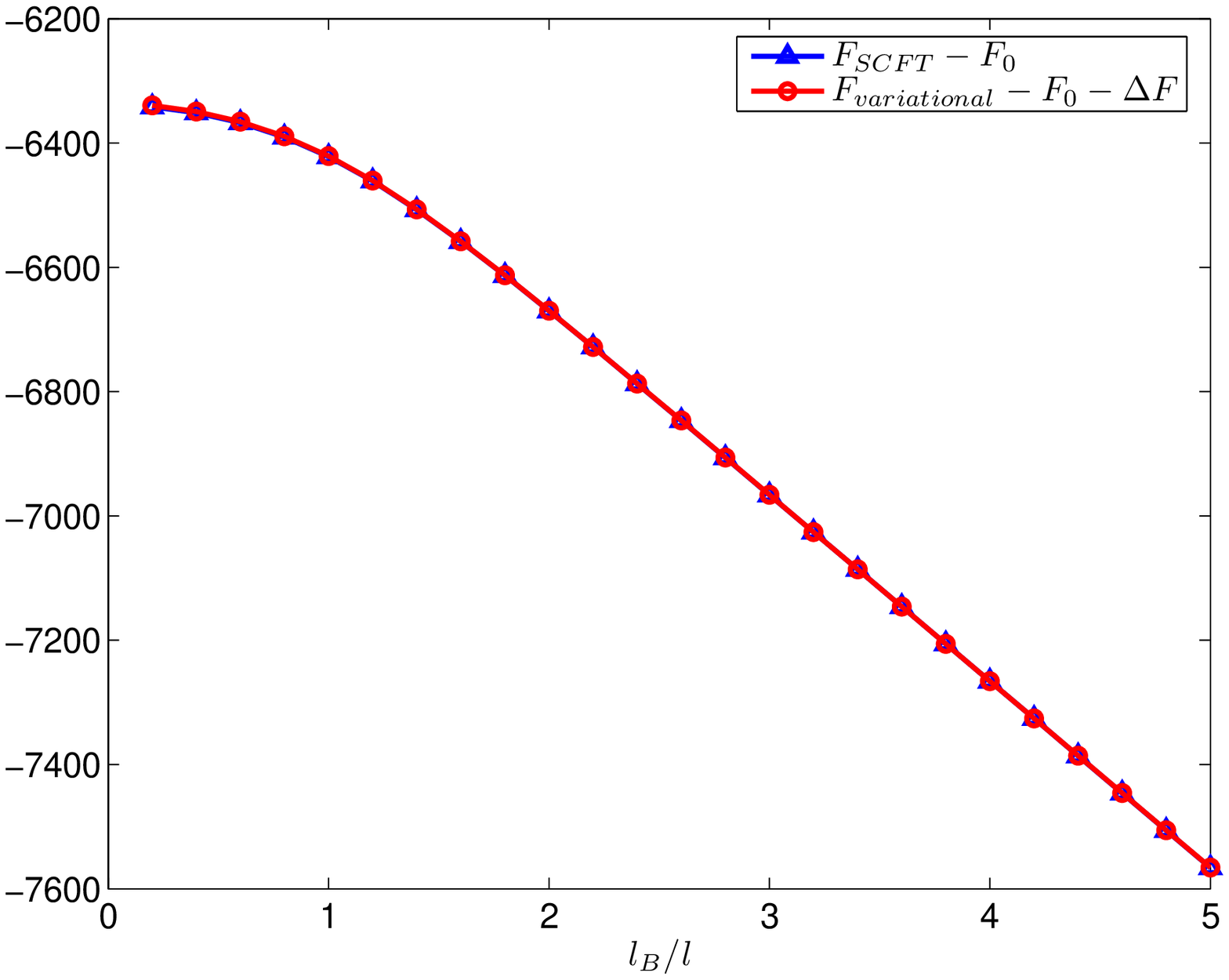}
    \end{minipage}
\caption{} \label{fig:compare_free_all}

\end{center}
\end{figure}

\begin{figure}[h!]

\begin{center}
$\begin{array}{c@{\hspace{.1in}}c@{\hspace{.1in}}}
\includegraphics[width=3in,height=3in]{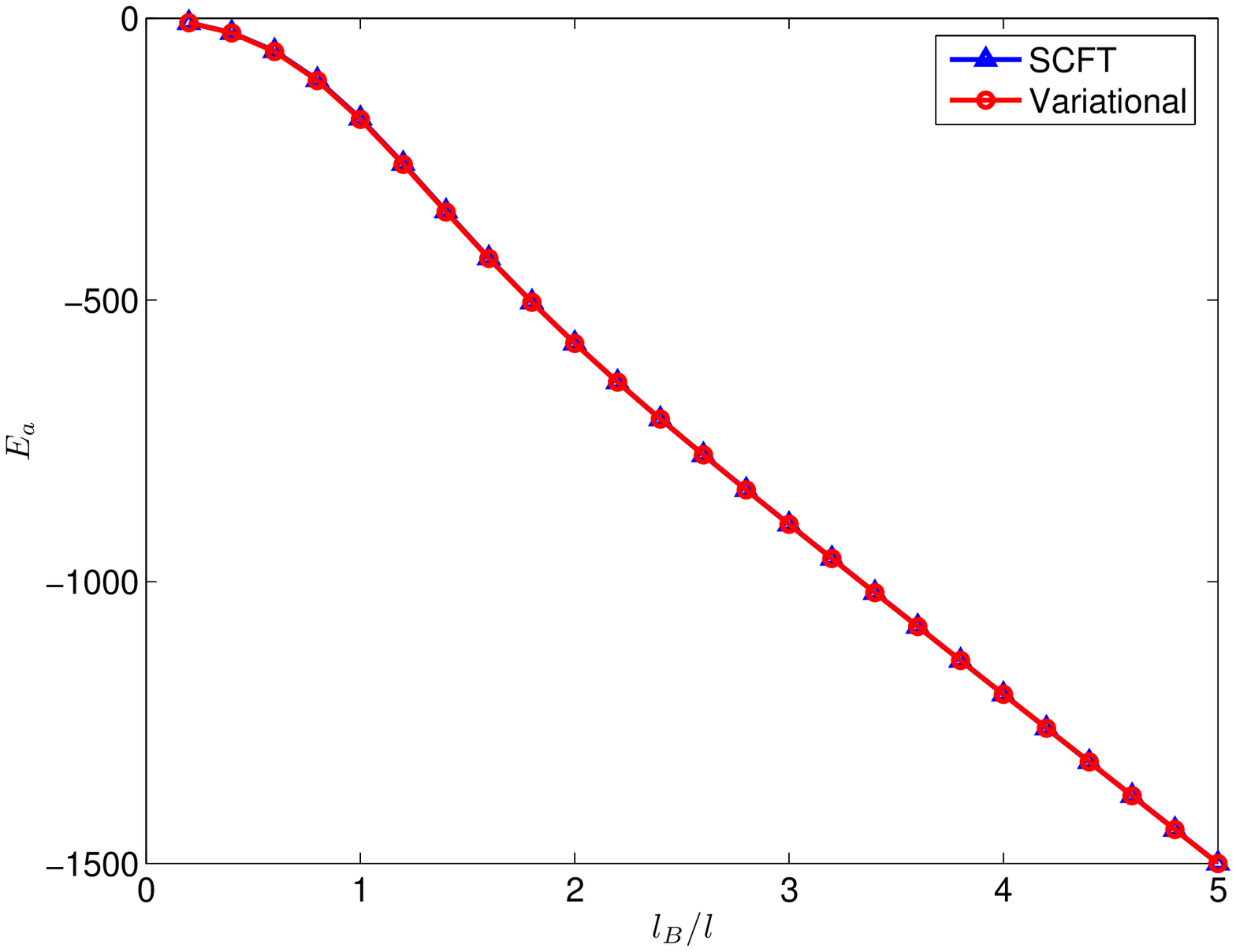}  &
\includegraphics[width=3in,height=3in]{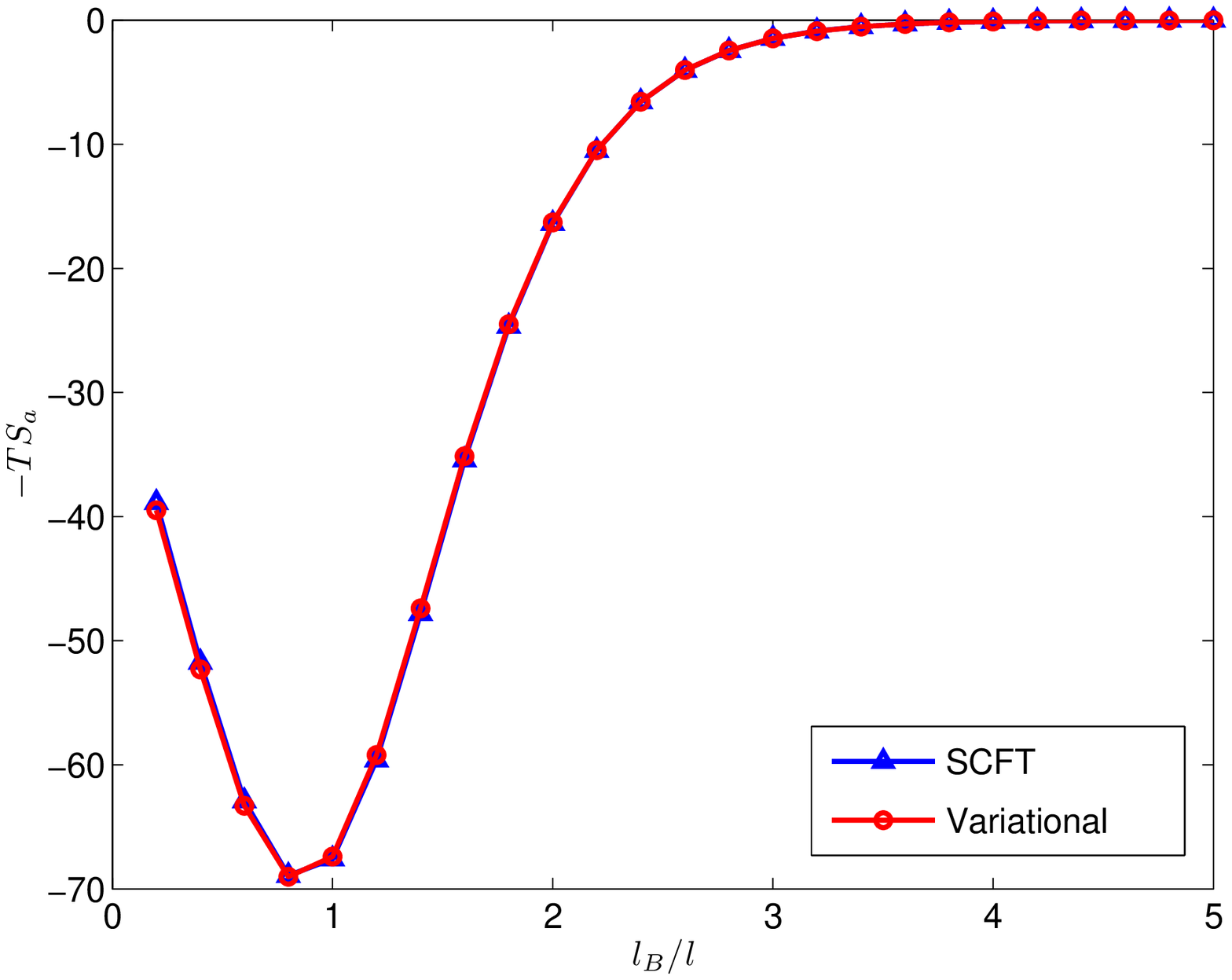} \\
\textbf{(a)} & \textbf{(b)}
\end{array}$

$\begin{array}{c@{\hspace{.1in}}c@{\hspace{.1in}}}
\includegraphics[width=3in,height=3in]{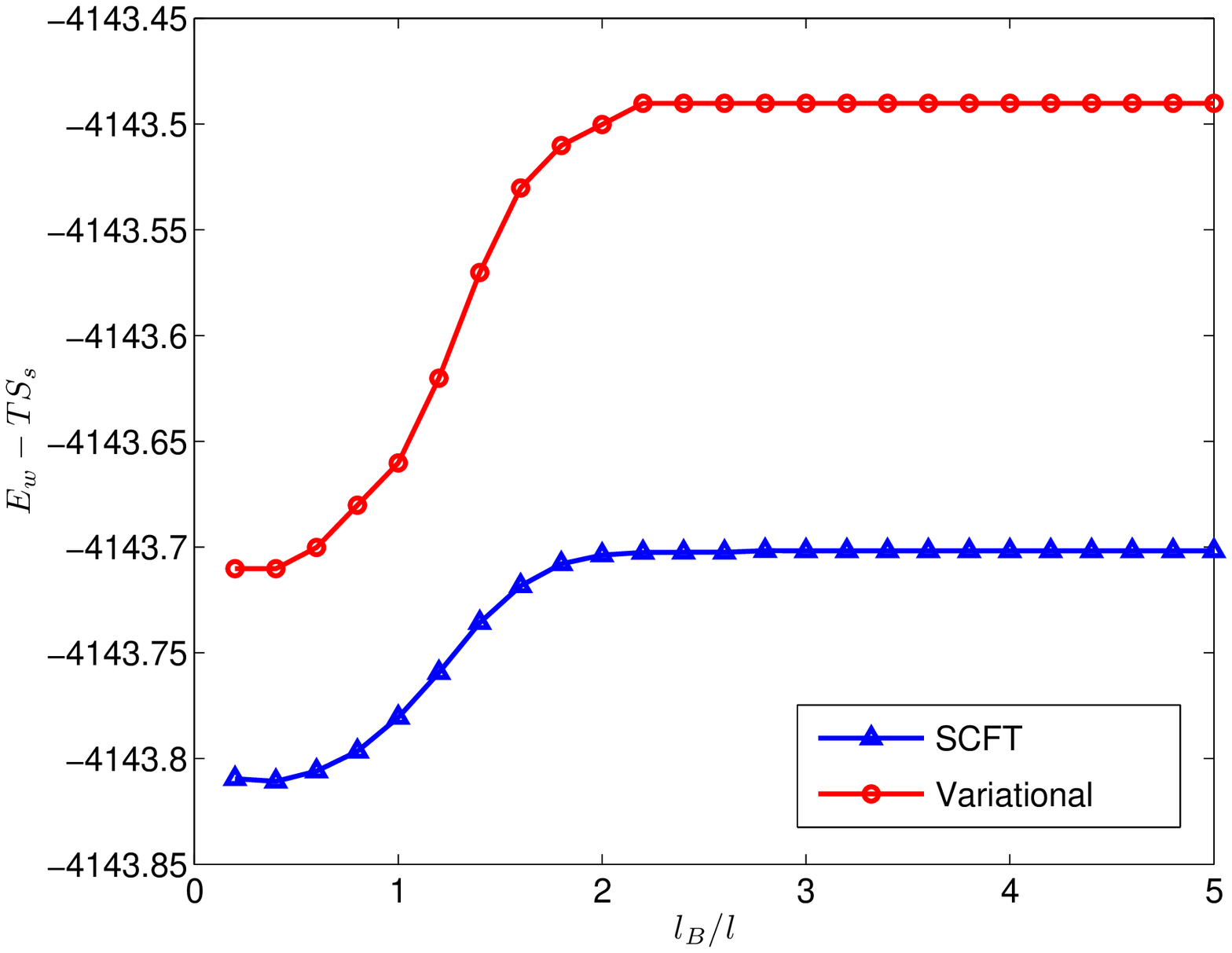}  &
\includegraphics[width=3in,height=3in]{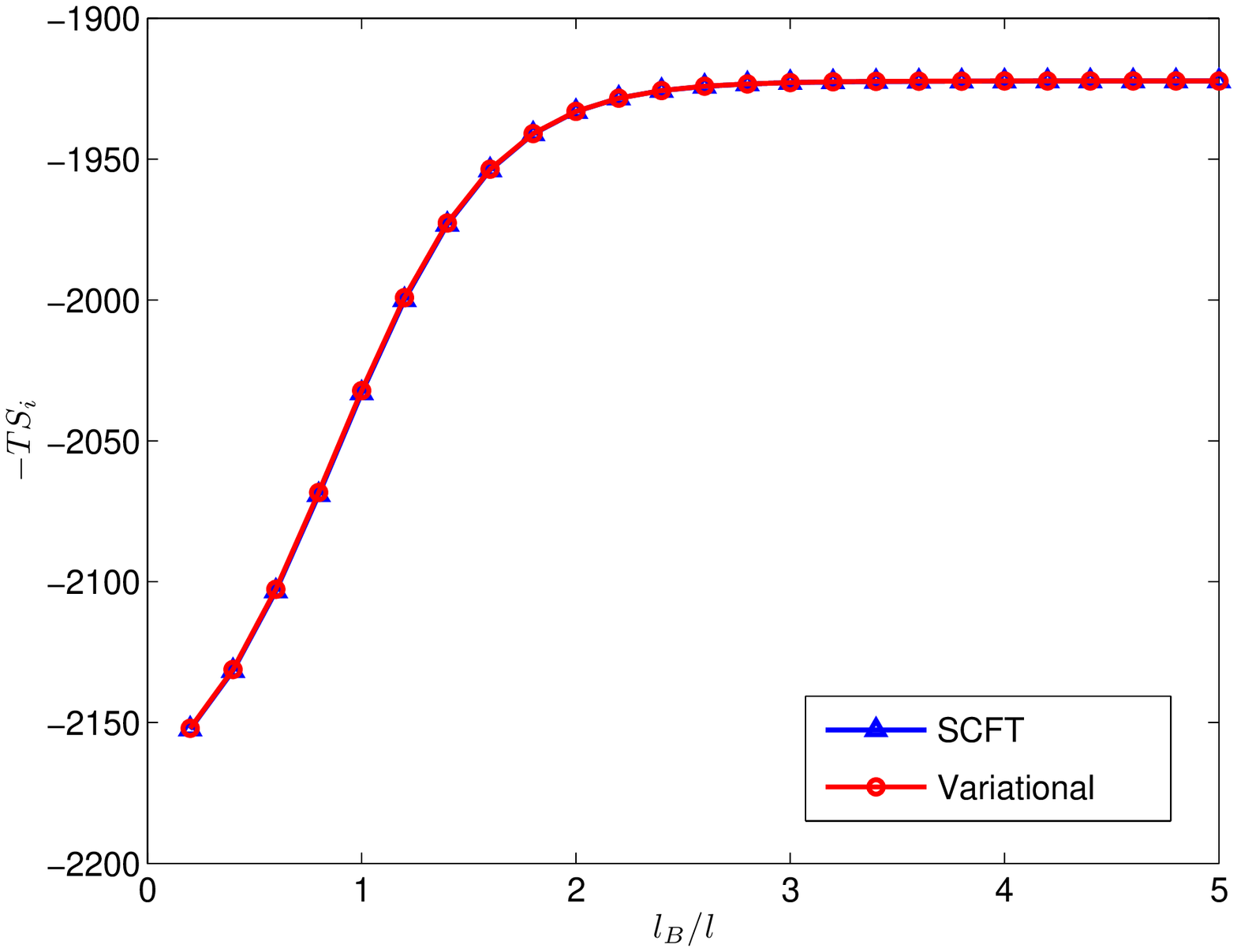}\\
\textbf{(c)} & \textbf{(d)}
\end{array}$

\end{center}
\caption{} \label{fig:compare_free_major}
\end{figure}

\begin{figure}[h!]
\begin{center}
$\begin{array}{c@{\hspace{.1in}}c@{\hspace{.1in}}}
\includegraphics[width=3.8in,height=3.8in]{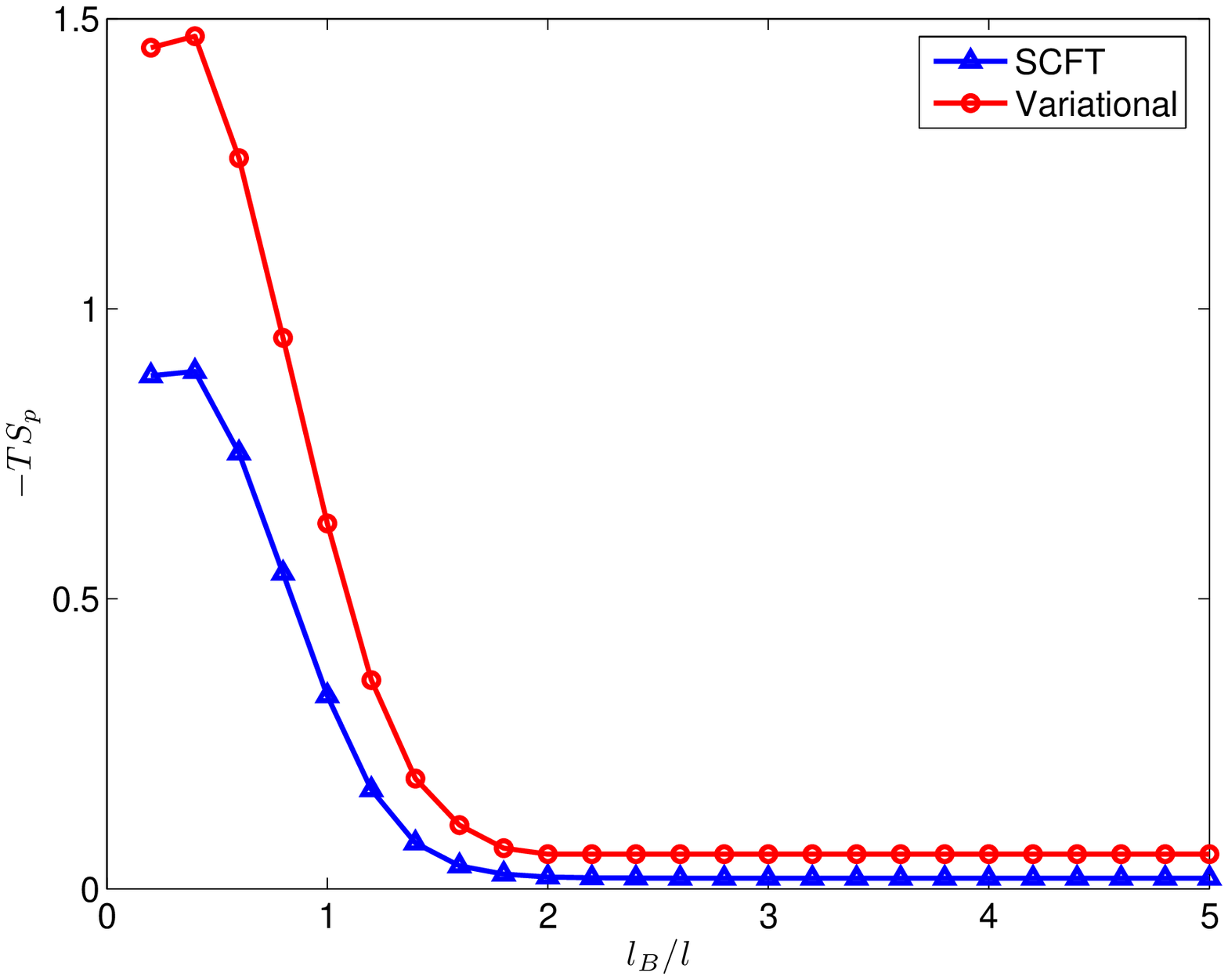} &  \\
\textbf{(a)} & \\
\includegraphics[width=3.8in,height=3.8in]{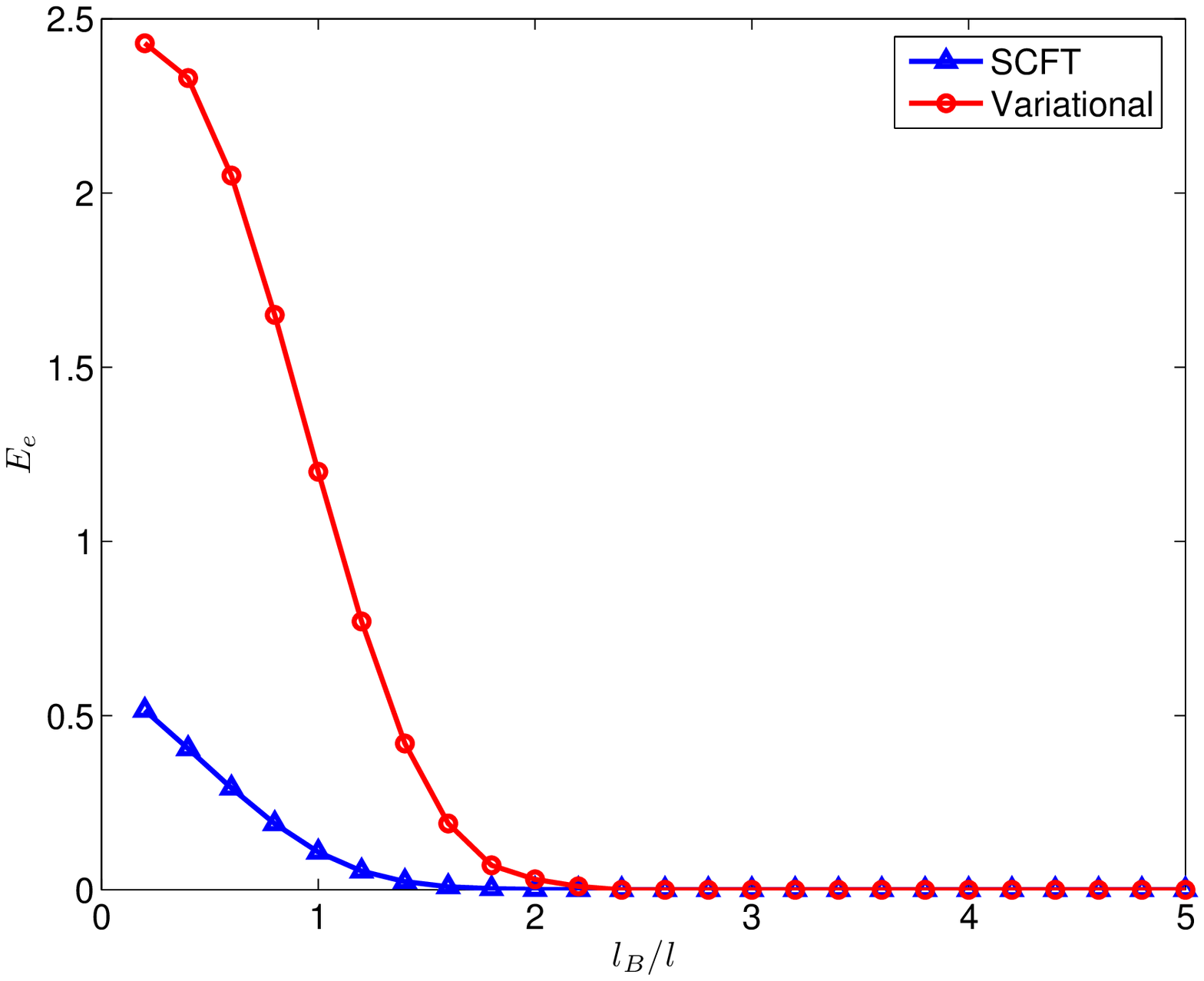} & \\
 \textbf{(b)} &
\end{array}$
\end{center}
\caption{} \label{fig:compare_free_minor}
\end{figure}

\begin{figure}[h]
\begin{center}
     \vspace*{1.0cm}
      \begin{minipage}[c]{15cm}
     \includegraphics[width=15cm]{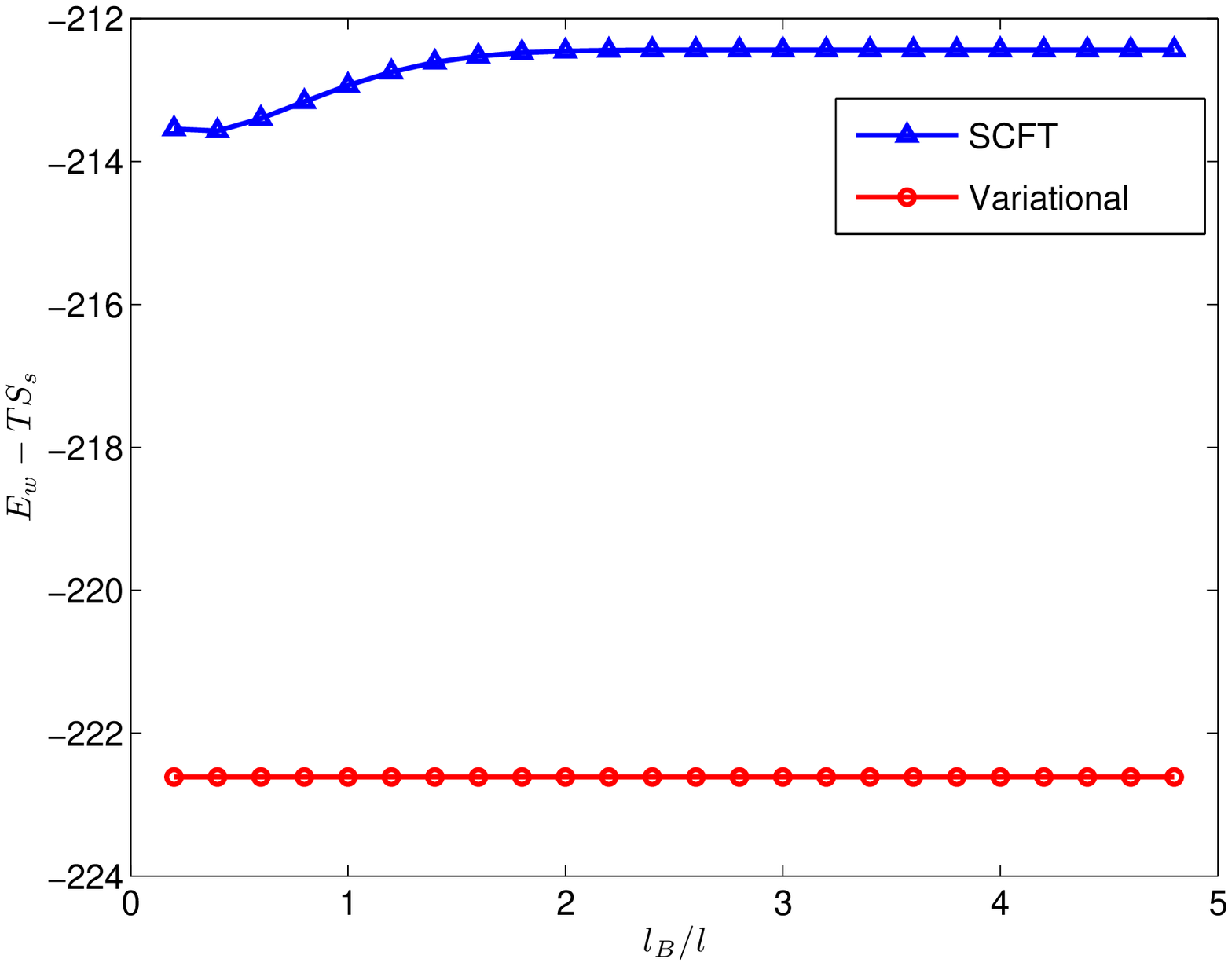}  
    \end{minipage}
\caption{}\label{fig:compare_ew_tss_high_density}
\end{center}
\end{figure}

\begin{description}
\item[Fig. 4.:] Comparison of major contributions to the free energies (presented in Fig. ~\ref{fig:compare_free_all})
obtained from SCFT and the variational formalism. (a) Ion-pair energy contributions ($E_a$), (b) translational entropy of
the ``adsorbed'' counterions ($-TS_a$), (c) polymer-solvent interaction energy and solvent entropy($E_w - TS_s$); and (d)
translational entropy of the ``free'' ions ($-TS_i$). The variational theory captures SCFT results quantitatively.
\end{description}

\begin{description}
\item[Fig. 5.:] Comparison of minor contributions to the free energies (presented in Fig. ~\ref{fig:compare_free_all}). (a)
Conformational entropy of the chain ($-TS_p$) and (b) electrostatic energy ($E_e$).
\end{description}

\begin{description}
\item[Fig. 6.:] A discrepancy between SCFT and variational theory for the polymer-solvent interaction energy and solvent entropy contributions arise
at high monomer densities. $Z_p = -Z_c = -1, R/l = 4, N = 100, c_{s} = 0.1M, \chi_{ps} = 0.45$ and $\delta = 3$.
\end{description}

\newpage
\vspace*{1.0cm}

\newpage
\vspace*{1.0cm}

\newpage
\vspace*{1.0cm}

\newpage

\newpage

\newpage
\vspace*{1.0cm}

\newpage
\begin{center}
{\bf{For Table of Contents Use Only}}\\
Counterion adsorption on flexible polyelectrolytes: comparison of theories\\
Rajeev Kumar, Arindam Kundagrami and M.Muthukumar \\
\it{Dept. of Polymer Science \& Engineering,\\
 Materials Research Science \& Engineering Center,\\
 University of Massachusetts,
 Amherst, MA-01003, USA.}
\end{center}
\vspace*{1.0cm}
\begin{figure}[h]
\begin{center}
          \includegraphics[width=9cm,height=3.75cm]{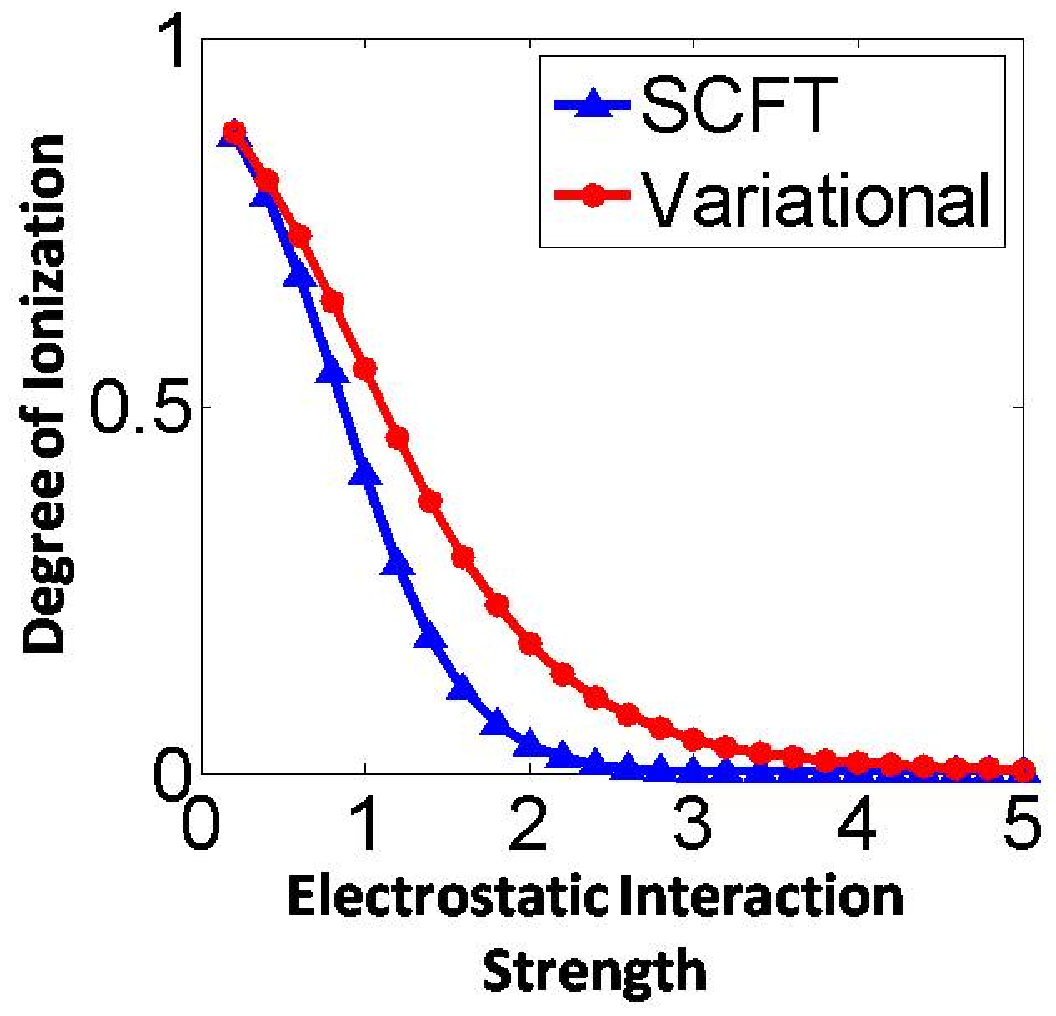}  
 
\end{center}
\end{figure}

\end{document}